\documentstyle[twocolumn,aps,epsfig,floats]{revtex}

\begin{document}

\twocolumn[\hsize\textwidth\columnwidth\hsize\csname %
@twocolumnfalse\endcsname

\title {Resonant Impurity States in the D-Density-Wave Phase}
\author{Dirk K.~Morr}
\address{
Theoretical Division, Los Alamos National Laboratory, Los Alamos,
NM 87545}
\date{\today}
\draft \maketitle
\begin{abstract}
We study the electronic structure near impurities in the
d-density-wave (DDW) state, a possible candidate phase for the
pseudo-gap region of the high-temperature superconductors. We show
that the local DOS near a non-magnetic impurity in the DDW state
is {\it qualitatively} different from that in a superconductor
with $d_{x^2-y^2}$-symmetry. Since this result is a robust feature
of the DDW phase, it can help to identify the nature of the two
different phases recently observed by scanning tunneling
microscopy experiments in the superconducting state of underdoped
Bi-2212 compounds.
\end{abstract}
 ]

\narrowtext

The pseudogap region of the high-temperature superconductors
(HTSC) has attracted considerable theoretical and experimental
attention over the last few years \cite{PG}. Recent scanning
tunneling microscopy (STM) experiments~\cite{STM} in the
superconducting state of underdoped Bi-2212 compounds have
observed regions with two {\it qualitatively} different density of
states (DOS). These regions are referred to as the $\alpha$ and
$\beta$ phase. While the DOS in the $\alpha$ phase resembles that
of a superconductor with $d_{x^2-y^2}$-symmetry, the DOS in the
$\beta$ phase is similar to that observed in the pseudogap phase
of the underdoped HTSC. A series of papers \cite{DDW,Cha01} have
proposed that the unusual phenomenology of the pseudogap region
arises from the presence of a d-density-wave (DDW) phase; a
current-carrying state with $d_{x^2-y^2}$-symmetry. Whether the
$\beta$ phase can be identified with the DDW state is currently a
topic of intense discussion.

In this Letter we suggest that the nature of the observed phases
can be clarified by considering the electronic structure in the
vicinity of impurities. We show, using a $\hat{T}$-matrix
approach, that the local DOS near a non-magnetic impurity in the
DDW state is {\it qualitatively} different from that in a
superconductor with $d_{x^2-y^2}$-symmetry (dSC). While the
DDW-DOS near an impurity changes with the specific form of the
normal state band structure, its qualitative differences to the
DOS in a dSC remain. They are therefore robust features
of the DDW phase which can they shed important insight into and
discriminate between the $\alpha$ and $\beta$ phases observed in
Refs.~\cite{STM}. Finally, we show that a magnetic impurity induces
two resonance
states in the DDW-DOS, a result which is qualitatively similar to
that in a dSC.

Starting point for our calculations is the $T$-matrix formalism
for a single impurity in the DDW state. We introduce the spinor
\begin{equation}
\Psi^\dagger_{k,\alpha}=\left(c^\dagger_{k,\alpha},c^\dagger_{k+Q,\alpha}
\right) \ ,
\end{equation}
where $\alpha$ is the spin index, such that the electronic Greens function is
given by
$\hat{G}_{\alpha,\alpha}({\bf k},
\tau-\tau^\prime) =- \langle {\cal T} \Psi_{k,\alpha}(\tau)
\Psi^\dagger_{k,\alpha}(\tau^\prime) \rangle$. In what follows we omit the spin
index, since for the cases considered below $\hat{G}$ is independent of
$\alpha$. In
the clean limit one has
\begin{equation}
{\hat G}_0^{-1}({\bf k},\omega_n) = \left( \begin{array}{cc} i
\omega - \epsilon_k & i \Delta_k \\ -i \Delta_k & i \omega -
\epsilon_{k+Q}
\end{array}
\right)  \ ,
\end{equation}
where $\Delta_{\bf k}=\Delta_0(\cos k_x - \cos k_y)/2$ is
the DDW gap, ${\bf Q}=(\pi,\pi)$ is the ordering moment of the DDW
state. The normal state electronic dispersion is given by
\begin{equation}
\epsilon_{\bf k}=- 2 t ( \cos k_x+ \cos k_y) -4 t' \cos k_x \cos
k_y - \mu \ ,\label{nsdisp}
\end{equation}
where $t, t^\prime$ are the hopping elements between nearest and next-nearest
neighbors, respectively, and $\mu$ is the chemical potential. In the following
we use
$t=300$ meV, $\Delta_0=50$ meV, and values for $t^\prime$ and
$\mu$ as discussed below.

\begin{figure} [t]
\begin{center}
\leavevmode
\epsfxsize=7.5cm
\epsffile{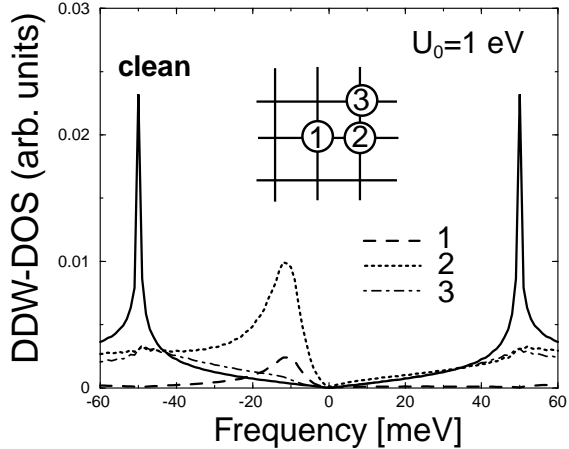}
\end{center}
\caption{DDW-DOS for the clean case
(solid line) and in the presence of a non-magnetic impurity with
$U_0=1$ eV: (1) DOS on the impurity site, (2) DOS on the
nearest-neighbor site, and (3) DOS on the next-nearest-neighbor
site. } \label{DOSDDW}
\end{figure}

We first consider a non-magnetic impurity described by the
scattering Hamiltonian
\begin{equation}
{\cal H}_{sc}=U_0 {\sum_{{\bf k,k^\prime},\alpha}}^{\hspace{-0.1cm}\prime}
\Psi^\dagger_{{\bf k},\alpha} \, \hat{\tau} \, \Psi_{{\bf k}^\prime,\alpha} \ ,
\end{equation}
where $U_0$ is the scattering strength of the impurity, the prime
restricts the summation to the momenta in the Brillouin zone (BZ)
of the DDW state, $\hat{\tau}=\hat{1}+\hat{\tau}_2$, $\hat{1}$ is
the unit matrix and $\hat{\tau}_{\alpha}$ are the Pauli matrices
in spinor space. In the presence of this scattering potential the
Greens function is given by
\begin{eqnarray}
\hat{G}({\bf r},{\bf r}',i \omega_n) &=& \hat{G}_0({\bf
r}^\prime-{\bf r},i\omega_n) + \nonumber \\ & & \hspace{-1.0cm}
\hat{G}_0(-{\bf r},i\omega_n) \hat{T}(i\omega_n) \hat{G}_0({\bf
r}',i\omega_n) \ , \label{fullG}
\end{eqnarray}
where one has for the $\hat{T}$-matrix
\begin{equation}
\hat{T}(i\omega_n) = [\hat{1}-U_0 \, \hat{\tau} \, \hat{G}_0(0,
i\omega_n)]^{-1} \ U_0 \hat{\tau} \ .
\label{Tmatrix1}
\end{equation}
For the change in the electron Greens function due to the impurity
scattering one obtains
\begin{equation}
\delta \hat{G}^{(11)}({\bf r}, i\omega_n)= U_0{\left[
\hat{G}^{(11)}_0({\bf r}, i\omega_n)\right]^2+\left[
\hat{G}^{(12)}_0({\bf r}, i\omega_n)\right]^2 \over 1-U_0 \,
\hat{G}^{(11)}_0(0, i\omega_n)} \ . \label{dG}
\end{equation}
Note, that for a given scattering strength, $U_0$, there exist
only a {\it single} resonant state at a frequency, $\omega_{res}$,
where the real part of the denominator on the r.h.s.~of
Eq.(\ref{dG}) vanishes. We first consider an electronic structure
in the normal state, Eq.(\ref{nsdisp}), with $t^\prime=0$ and at
half-filling, $\mu=0$. We present our numerical results for the
DDW-DOS
\begin{equation}
N(\omega)=-{2 \over \pi} \ {\rm Im} \,
\hat{G}^{(11)}({\bf r,r}, \omega+i\delta) \ ,
\end{equation}
in the absence of an impurity (clean case) and for a single
non-magnetic impurity with scattering strength $U_0=1$ eV in
Fig.~\ref{DOSDDW}.
The DDW-DOS in the clean case (solid line) vanishes linearly at
small frequencies, and exhibits two peaks at $\pm \Delta_0$.  As
discussed above, a {\it single} resonance state appears at the
impurity site (dashed line). For $U_0>0 \, (<0)$, this resonant
state is located at $\omega_{res}<0 \, (>0)$, i.e., on the
particle (hole) site. In the DOS on the impurity's
nearest-neighbor site (dotted line) the resonant state remains
particle-like with an amplitude which is larger than that on the
impurity site itself. This is to be expected since at the impurity
site, $\hat{G}^{(12)}_0=0$, while at the nearest neighbor site
$\hat{G}^{(12)}_0\not =0$. In the limit $U_0 \rightarrow \infty$,
the spectral weight of the resonance at the impurity site, $U_0
[\hat{G}^{(11)}_0(0, \omega_{res})]^2$ vanishes. The DOS on the
impurity's next-nearest-neighbor site exhibits barely any resonant
enhancement, while the peaks at the gap edges are still
suppressed. Thus, as one moves away from the impurity site, the
signature of the resonance states in the DOS are lost before the
gap edge peaks are recovered.

\begin{figure} [t]
\begin{center}
\leavevmode
\epsfxsize=7.5cm
\epsffile{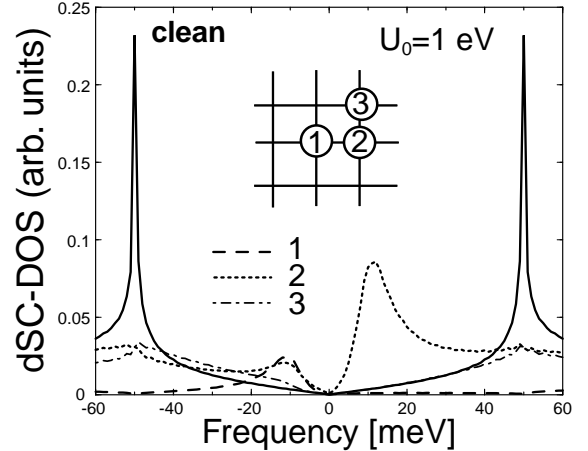}
\end{center}
\caption{dSC-DOS for the clean case (solid line) and in the
presence of a non-magnetic impurity with $U_0=1$ eV: (1) DOS on
the impurity site, (2) DOS on the nearest-neighbor site, and (3)
DOS on the next-nearest-neighbor site.} \label{DOSSC}
\end{figure}

We now compare the above results for the DOS with those near a
non-magnetic impurity in a superconductor with
$d_{x^2-y^2}$-symmetry. Here, the correction to the Greens
function arising from the impurity scattering is given by
\cite{Shiba68,Martin,Sachdev,Flatte}
\begin{equation}
\delta G(r,i\omega_n)={U_0 \left[ G_0({\bf r}, i\omega_n)\right]^2
\over 1-U_0 \, G_0(0, i\omega_n)} - {U_0 \left[ F_0({\bf r},
i\omega_n)\right]^2 \over 1-U_0 \, G_0(0, -i\omega_n)} \ ,
\label{dGsc}
\end{equation}
where $G_0,F_0$ are the normal and anomalous Greens functions in the
SC state, respectively. Due to the particle-hole mixing in the SC
state, there exist in general two resonance states at frequencies
where the denominators on the r.h.s.~of Eq.(\ref{dGsc}) vanish.
Note, that the spectral weight of these states is determined by
$\left[ G_0({\bf r}, i\omega_n)\right]^2, \left[ F_0({\bf r},
i\omega_n)\right]^2$, respectively. In Fig.~\ref{DOSSC} we present
the DOS near the impurity site for $U_0=1$ eV, and the same set of
band parameters as in Fig.~\ref{DOSDDW}.
The SC DOS in the clean limit (solid line) is identical to that in
the DDW state (solid line in Fig.~\ref{DOSDDW}). The DOS at
the impurity site exhibits a {\it single} resonance state, since
the spectral weight of the second resonance vanishes,
$\left[ F_0(0, i\omega_n)\right]^2 \equiv 0$. For $U_0>0 \, (<0)$
the resonance state is located at $\omega_{res}<0 \, (>0)$, i.e.,
on the particle (hole) site. While this result is similar to that
in the DDW state (dashed line in Fig.~\ref{DOSDDW}), the DOS on
the impurity's nearest-neighbor site is {\it qualitatively}
different (dotted line). Since here the spectral weight of
both resonances is non-zero, $\left[ G_0({\bf r},
i\omega_n)\right]^2, \left[ F_0({\bf r}, i\omega_n)\right]^2 \not
= 0$, the DOS exhibits two resonance states. We find that in
general, i.e., independent of the specific band structure, the
hole-like resonance possesses the larger spectral weight,
$\left[ F_0({\bf r}, i\omega_n)\right]^2 > \left[ G_0({\bf r},
i\omega_n)\right]^2$. Similar results were obtained earlier in
Refs.~\cite{Martin,Sachdev}. Thus, the spatial dependence of the
DOS near a non-magnetic impurity can be used to determine the
broken-symmetry phase in which the impurity is embedded. A similar
conclusion was recently reached by Zhu {\it et al.}~\cite{Zhu01}.
Note, that the appearance of a resonance state at the impurity
site only requires a suppression of the DOS at low frequencies
\cite{Kruis}, but not necessarily a broken symmetry phase.

\begin{figure} [t]
\begin{center}
\leavevmode
\epsfxsize=7.5cm
\epsffile{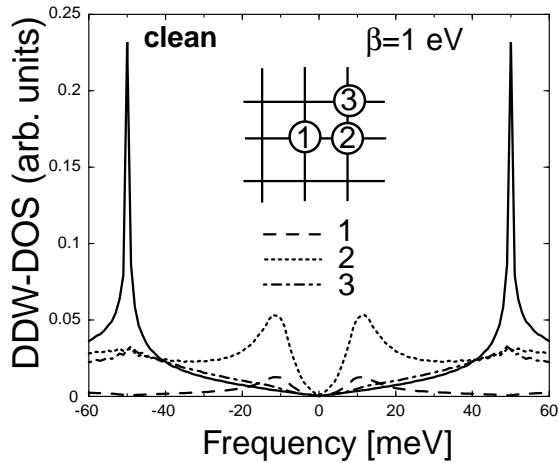}
\end{center}
\caption{DDW-DOS for the clean case
(solid line) and in the presence of a magnetic impurity with
$\beta=1$ eV:  (1) DOS on the impurity site, (2) DOS on the
nearest-neighbor site, and (3) DOS on the next-nearest-neighbor
site. } \label{DOSDDWmag}
\end{figure}

We next consider the electronic states in the vicinity of a
magnetic impurity described by the scattering Hamiltonian
\begin{equation}
{\cal H}_{sc}=-JS {\sum_{{\bf
k,k^\prime},\alpha,\beta}}^{\hspace{-0.25cm}\prime}
\Psi^\dagger_{{\bf k},\alpha} \, \hat{\tau} \, \hat{\sigma}_{\alpha,\beta}
\, \Psi_{{\bf k}^\prime,\beta}
\end{equation}
where $S=1/2$ and $\hat{\sigma}$ are the Pauli-matrices in spin
space. For the $\hat{T}$-matrix one has \cite{com1}
\begin{equation}
\hat{T}(i\omega_n) = \left[\hat{1}-\left(\beta \, \hat{\tau} \,
\hat{G}_0(0, i\omega_n)\right)^2 \right]^{-1} \beta^2 \hat{\tau}
\hat{G}_0(0, i\omega_n) \hat{\tau} \label{Tmatrix2}
\end{equation}
with $\beta=JS$. The correction to the Greens function arising
from the impurity scattering is given by
\begin{eqnarray}
\delta \hat{G}^{(11)}({\bf r}, i\omega_n) &=& { \beta^2
\hat{G}^{(11)}_0(0, \omega) \over 1-\beta^2 \,
\left[\hat{G}^{(11)}_0(0, \omega)\right]^2} \nonumber \\ & &
\hspace{-1.5cm} \times \left\{ \left[ \hat{G}^{(11)}_0({\bf r},
i\omega_n) \right]^2+\left[ \hat{G}^{(12)}_0({\bf r},
i\omega_n)\right]^2 \right\} \ . \label{dGm}
\end{eqnarray}
Note, that a magnetic impurity in the DDW state induces two
resonance states with {\it equal} spectral weight. In
Fig.~\ref{DOSDDWmag} we present the DOS in the clean case and for
a single magnetic impurity with scattering strength $\beta=1$ eV.
The resonance states again posses a larger spectral weight at the
nearest-neighbor site than on the impurity site itself. Similar to
the case of a non-magnetic impurity, the DOS on the
next-nearest-neighbor site exhibits barely any resonant
enhancement, while the peaks at the gap edges are still
suppressed. These results are qualitatively similar to those
obtained for a magnetic impurity in a dSC.

\begin{figure} [t]
\begin{center}
\leavevmode
\epsfxsize=7.5cm
\epsffile{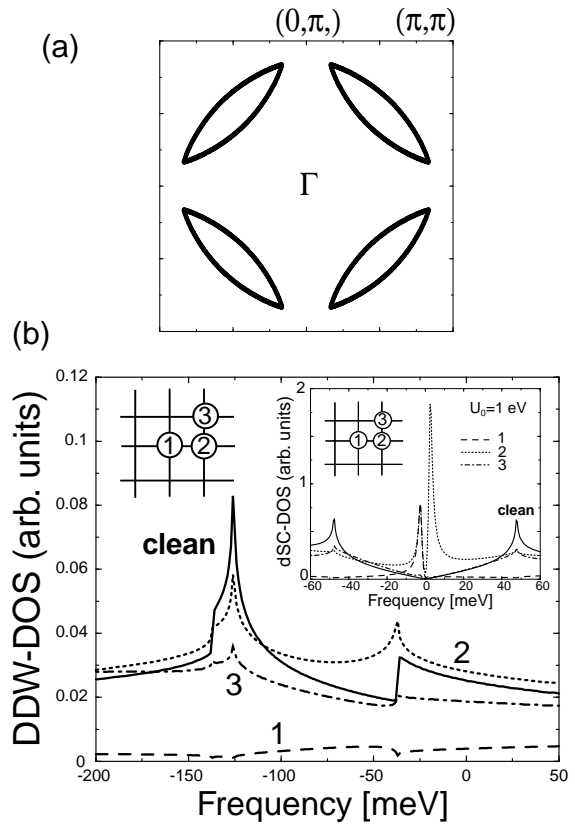}
\end{center}
\caption{{\it (a)} Fermi surface in the DDW state with
$t^\prime=-0.3t$, $\mu=-0.91t$ (corresponding to a hole-doping of
$10\%$) and $\Delta_0=50$ meV. The hole pockets are centered
around $(\pm \pi/2,\pm \pi/2)$. {\it (b)} DOS in the DDW state
with the same band parameters as in {\it (a)}, for the clean case
(solid line) and in the presence of a non-magnetic impurity with
$U_0=1$ eV: (1) DOS on the impurity site, (2) DOS on the
nearest-neighbor site, and (3) DOS on the next-nearest-neighbor
site. Inset: SC DOS for the same band parameters as in {\it (a)}.
} \label{Fig4}
\end{figure}

So far, we have only discussed the DOS near an impurity for the
half-filled case and $t^\prime=0$. We next address the question of
how robust the above results are against changes in the band
structure. In particular, we consider a set of band parameters which
describes the FS measured by ARPES experiments in Bi-2212
\cite{ARPES}. To this end we take $t^\prime=-0.3t$ and
$\mu=-0.91t$, which corresponds to a hole doping of $10\%$,
characteristic of the underdoped HTSC. The resulting Fermi surface
in the DDW state is shown in Fig.~\ref{Fig4}a.
In contrast to the case considered above with $t^\prime=\mu=0$,
where the FS only consists of four Fermi points at ${\bf
q_c}=(\pi/2,\pm \pi/2)$, the FS now exhibits hole pockets centered
around ${\bf q_c}$. This form of the FS immediately implies that
the DOS (in the clean case) at zero energy is now non-zero, in
contrast to the results shown in Fig.~\ref{DOSDDW}. This is
confirmed by our numerical evaluation of the DOS which we present
in Fig.~\ref{Fig4}b. In the clean case (solid line) the DOS
exhibits two peaks at $\omega \approx -126$ and $-35$ meV, which
correspond to the two peaks at $\omega=\pm 50$ meV for
$t^\prime=\mu=0$ (see Fig.~\ref{DOSDDW}). Thus, the DOS for non-zero
$t^\prime$ and hole-doping is shifted downwards in energy, in
comparison to the case $t^\prime=\mu=0$. Since the DOS in
Fig.~\ref{Fig4}b does {\it not} show any significant reduction at
low energies, we expect that any resonance state near an impurity
couples strongly to the continuum of electronic states, which in
turn reduces its amplitude and spectral weight. Our numerical
results for the DOS near a non-magnetic impurity with $U_0=1$ eV
as shown in Fig.~\ref{Fig4}b confirm this conclusion. While the
DOS at the impurity site is again reduced in comparison to the
clean case, it does not exhibit any signature of a resonance
state. In contrast, the DOS in the SC state remains qualitatively
unchanged from the case $t^\prime=\mu=0$, as shown in the inset of
Fig.~\ref{Fig4}b. We obtain {\it quantitatively} similar results
to those shown in Fig.~\ref{Fig4}b for a variety of doping
concentrations and values of $t^\prime$. We thus conclude that the
DOS in the DDW phase remains {\it qualitatively} different from
that in a dSC for a wide range of electronic band structures and chemical
potentials. Our results are
therefore robust features of the DDW phase and can discriminate it
from other broken-symmetry phases.

The clean DOS in the DDW state for $t^\prime=\mu=0$ (Fig.~\ref{DOSDDW}) is in
good qualitative agreement with the results presented in Refs.~\cite{STM}. The
DDW-DOS for $t^\prime=-0.3t$ and $10 \% $ hole doping (Fig.~\ref{Fig4}b),
however, does not show a
reduction at low frequencies, and thus apparently disagrees with the
experimental data. Here, we propose two
explanations to resolve this apparent discrepancy. First, the differences
between the $\alpha$ and $\beta$ phases could arise from doping
inhomogeneities on the $20-30 \AA$ scale, with the $\alpha (\beta)$ phase
containing the higher (lower) hole concentration. According to the phase
diagram proposed in Ref.~\cite{Cha01}, the $\beta$ phase then corresponds to the
DDW state, while the $\alpha$ phase is a dSC. In this case,
the DDW-DOS is expected to be similar to that shown in Figs.~\ref{DOSDDW}
and \ref{DOSDDWmag} (small $\mu$ limit), while the dSC-DOS resembles our results
in the inset of Fig.~\ref{Fig4}b (large $\mu$ limit), in qualitative agreement
with
the
experimental data.
Second, it was argued that the interaction with the spin susceptibility,
$\chi$, peaked at ${\bf
Q}=(\pi,\pi)$, substantially changes the properties of
electrons in those regions of
the FS which can be connected by ${\bf Q}$ (hot
spots) \cite{Pines,Chu}.
In particular, it was shown that for the underdoped
HTSC, this
interaction leads to a suppression of the
DOS at low frequencies and a shift of spectral weight to higher
energies \cite{Pines}. Since the FS in the DDW state (see
Fig.~\ref{Fig4}a) exhibits a high degree of nesting with
wave-vector ${\bf Q}$, we expect that the interaction with spin
excitations yields the same kind of suppression in the
low frequency DDW-DOS. This suppression then leads
to a reappearance of
resonance states near the impurity site.
Work to explore this possibility is
currently under
way \cite{Kos01}. If this conjecture turns out to be correct,
the
suppression of the low energy DDW-DOS would be a
robust feature and
independent of the particular form of the FS in
the normal state.

In conclusion, we have shown that a non-magnetic impurity in the
DDW state
induces a resonance state in the local DOS. At half-filling and $t^\prime=0$ we
argue that the
spatial dependence of this resonance peak is {\it qualitatively}
different from that in
a superconductor with $d_{x^2-y^2}$-symmetry. Since away
from half-filling, the qualitative differences between the
DDW-DOS and the
dSC-DOS remain, they are robust features
of the DDW phase and can thus shed
important insight into the nature of the $\alpha$ and $\beta$ phases observed in
underdoped Bi-2212 \cite{STM}. Finally, we show that a magnetic
impurity induces two resonance
states in the DDW-DOS, a result which is
qualitatively similar to that in a dSC.

We would like to thank A. Balatsky, J.C. Davis, and D. Pines for
stimulating discussions, and C. Nayak for helpful comments. This
work has been supported by the Department of Energy at Los Alamos.


\begin{thebibliography}{99}
\bibitem{com1}
Similar to the approach by Shiba \protect\cite{Shiba68} we average
over the spin orientation of the impurity and thus only consider
scattering processes which are even in $(J {\bf S})$.
\end{thebibliography}
\end{document}